\documentclass[twocolumn,final,twoside,journal]{IEEEtran}

\usepackage{amsmath,amsthm,graphicx,bm}
\usepackage{epsfig,amsfonts}
\usepackage{graphicx}
\usepackage{color}
\usepackage[nolist]{acronym}
\usepackage{psfrag}
\usepackage{perso}
\usepackage{xspace}

\usepackage{pgfplots}
 \pgfplotsset{compat=newest}
    \pgfplotsset{plot coordinates/math parser=false}
    \pgfplotsset{
    label style={anchor=near ticklabel},
    xlabel style={yshift=0.0em},
    ylabel style={yshift=-0.3em},
    tick label style={font=\footnotesize },
    label style={font=\footnotesize},
    legend style={font=\footnotesize},
    title style={font=\fontsize{7}}}

\usepackage{mathtools}
\mathtoolsset{showonlyrefs}

\usepackage{placeins}

\usepackage{multirow}
\usepackage{etoolbox}

\usepackage{flushend}
\usepackage{placeins} 

\usepackage{algorithm}
\usepackage[noend]{algpseudocode}

\usepackage{subfig}
\usepackage[normalem]{ulem}
\usepackage{multicol}

\usepackage{amsthm}
\newtheorem{example}{Example}
\usepackage{booktabs}
\usepackage{booktabs}
\usepackage{ulem}

\renewcommand{\S}{\mathcal{S}}
\newcommand{\SA}{\S_A}
\newcommand{\SB}{\S_B}

\newcommand{\x}{\mathrm{x}}

\newcommand{\countfield}{\emph{count}\xspace}
\newcommand{\payload}{\emph{data}\xspace}

\newcommand{\hashlength}{\nu}
\newcommand{\keylength}{\kappa}

\newcommand{\hash}{H_{m,d}}

\newcommand{\hashdegree}{h_{\mathbf{\Lambda}}}

\newcommand{\xor}{\text{XOR}\xspace}

\newcommand{\dmax}{{d_{\text{max}}}}
\newcommand{\jvn} {g} 
\newcommand{\jv} {\bm{\jvn}} 
\newcommand{\jvi}[1]{\jvn_{#1}}

\newcommand{\graph}{\mathcal{G}}
\newcommand{\cells}{\mathcal{C}}
\newcommand{\pairs}{\mathcal{Z}}
\newcommand{\edges}{\mathcal{E}}

\newcommand{\cell}{c}
\newcommand{\pair}{z}
\newcommand{\Prob}{\mathrm{P}}

\newcommand{\cellnode}{\mathtt{c}}
\newcommand{\pairnode}{\mathtt{z}}

\newcommand{\graphensemble}{\mathscr{G}(n,\eta,\lambda)}

\newcommand{\eff}{\eta}

\begin{document}
	
	\begin{acronym}
\acro{IBLT}{invertible Bloom lookup table}
\acro{IRSA}{irregular repetition coded slotted ALOHA}
\end{acronym}

\title{Irregular Invertible Bloom Look-Up Tables \vspace{-3mm}}

\author{ 
	Francisco L\'azaro, Bal\'azs Matuz \\
		Institute of Communications and Navigation of DLR (German Aerospace Center),
		\\Wessling, Germany. Email: \{Francisco.LazaroBlasco, Balazs.Matuz\}@dlr.de
	
\thanks{ This work has been accepted for presentation at the 11th International Symposium on Topics in Coding\\	
	\copyright 2021 IEEE. Personal use of this material is permitted. Permission
	from IEEE must be obtained for all other uses, in any current or future media, including
	reprinting /republishing this material for advertising or promotional purposes, creating new
	collective works, for resale or redistribution to servers or lists, or reuse of any copyrighted
	component of this work in other works}} 

\maketitle

\thispagestyle{empty} \pagestyle{empty}

	\begin{abstract}		
 	We consider \acp{IBLT} which are probabilistic data structures that allow to store key-value pairs. An \ac{IBLT} supports  insertion and deletion of key-value pairs, as well as the recovery of all key-value pairs that have been inserted, as long as the number of key-value pairs stored in the \ac{IBLT} does not exceed a certain number. The recovery operation on an \ac{IBLT} can be represented as a peeling process on a bipartite graph. We present a density evolution analysis of \acp{IBLT} which allows to predict the maximum number of key-value pairs that can be inserted in the table so that recovery is still successful with high probability. This analysis holds for arbitrary irregular degree distributions and generalizes results in the literature. We complement our analysis by numerical simulations of our own \ac{IBLT} design which allows to  recover a larger number of key-value pairs as state-of-the-art \acp{IBLT} of same size. 
	\end{abstract}

	\maketitle
	
	\setlength{\footskip}{12.3pt}
	

\vspace{-2mm}
\section{Introduction}\label{sec:Intro}
\Acfp{IBLT} were first introduced in \cite{goodrich:2011} as probabilistic data structures that can be used to represent a set $\S$ of elements. Every element of $\S$ is mapped to a  number $d$ of cells of the \ac{IBLT} (a formalization follows in Section~\ref{sec:Prelim}). We call an \ac{IBLT}  \textit{regular} if $d$ is a constant for every element of $\S$, otherwise, we say it is \textit{irregular}. In the context of data bases the elements of $\S$ are key-value pairs. 
The key can be thought of as a short (unique) identifier of an element in the database, whereas the value is the actual data which can be orders of magnitude larger than the key. Commonly, the key associated to an element of the database is obtained simply as a hash function of its value. 
As the name indicates, an important property of \acp{IBLT} is that they are invertible (in contrast to Bloom filters \cite{bloom1970space}), i.e., they allow to list the elements of the set $\S$ which they represent. 
The asymptotic performance of regular \acp{IBLT} was studied in \cite{goodrich:2011}. It was found that an \ac{IBLT} is invertible with high probability if its \textit{load}, defined as the ratio of key-value pairs to the number of cells, does not exceed the \emph{load threshold}. 
The analysis in \cite{goodrich:2011} relies on known results about the 2-core threshold of regular hypergraphs. In \cite{rink2013mixed}, the load threshold of specific irregular hypergraphs with only two different degrees was analyzed. 

In the literature, \acp{IBLT} are applied for so-called \textit{set reconciliation} problems aiming at establishing consistency among different sets of elements \cite{eppstein:2011}. In a two party system with sets $\SA$ and $\SB$ one would like to determine \textit{set differences} $\SA \setminus \SB$ and $\SB \setminus \SA$ in an efficient way and communicate the missing elements to the respective parties. 
Amongst others, \acp{IBLT} find applications in remote file synchronization, synchronisation of distributed databases, deduplication, or gossip protocols \cite{MitzenmacherP18,eppstein:2011}. Recently, \acp{IBLT} have been used to improve block propagation in the Bitcoin network \cite{Ozisik2019}. 

This work, extends the analysis of irregular \acp{IBLT}. We first illustrate that the recovery operation (sometimes also referred to as inversion) of an \ac{IBLT} corresponds to a peeling decoding process \cite{AEL95,luby1998analysis,luby:efficient} on a bipartite graph. Next, we derive a density evolution analysis  to obtain the load threshold. This generalizes the results of \cite{goodrich:2011,rink2013mixed} to arbitrary irregular \acp{IBLT}. Furthermore, we make the observation that the recovery process of \acp{IBLT} is strongly linked to the successive interference cancellation process for multiple access protocols over the collision channel \cite{Liva2011}. 
Finally, we provide an irregular \ac{IBLT} construction which outperforms the results in \cite{goodrich:2011,rink2013mixed}.

	\section{Irregular invertible Bloom lookup tables}\label{sec:Prelim}
\subsection{Description}
Let  $\S=\{\pair_1, \pair_2, ..., \pair_n\}$ be a set of elements, with $|\S|=n$. 
We assume that each element $\pair$ is a key-value pair, denoted by $\pair= (x,y)$. The key $x$ is of length $\hashlength$ bits and the value $y$ is of length $\keylength\gg \hashlength$ bits. The key $x$ is obtained as a function of $y$ where the mapping is many to one. 
For the analysis that follows we make two simplified, but common assumptions. 
First, all keys $x$ in the set are distinct, i.e., there are no key-collisions. Second,  the keys $x$ are selected uniformly from $\{0,1\}^{\hashlength}$. 

Let a cell $\cell$ be  a data structure containing two different fields \countfield and \payload where:
\begin{itemize}
    \item \countfield is an integer. It contains the number of elements that have been mapped to this cell (details on the mapping follow).
    \item \payload=(\payload.x, \payload.y) is a bit string of length $\hashlength+\keylength$ which can be divided into a pair of bit strings  of length $\hashlength$ and $\keylength$, respectively. The bit strings \payload.x and \payload.y contain, respectively, the  binary \xor of the keys and values that have been mapped to the cell.  
\end{itemize}

Let us define two hash functions:
\begin{itemize}
    \item $\hashdegree(x)=d$ is a non-uniform random hash function 
    which maps an input $x \in \{0,1\}^{\hashlength}$ to an output $d \in \{1,2,\dots,\dmax\}$. The parameter $\bm{ \Lambda} = ( \Lambda_1, \Lambda_2, \dots, \Lambda_{\dmax} )$, referred to as \textit{degree distribution}, is a probability mass function. Under the assumption that the input $x$ is uniformly distributed, we have  $P(d=i)=\Lambda_i$, i.e., the output of $\hashdegree(x)$ follows the degree distribution $\bm{ \Lambda}$.
    \item $\hash(x)=\jv$ is a random hash function which  maps an input $x \in \{0,1\}^{\hashlength}$ to a length-$d$ vector $\jv$ of $d$ \textit{different} natural numbers in $\{1,2,\ldots,m\}$, i.e., it samples $d$ different natural numbers between $1$ and $m$ \textit{without replacement}. Such a hash function can be obtained from a uniform random hash function that outputs a natural number between $1$ and $\prod_{i=0}^{d-1} (m-i)$. 
\end{itemize}

An irregular \ac{IBLT} is a probabilistic data structure to store elements of a set $\S$. It is defined by its degree distribution $\mathbf{\Lambda}$, the number of cells (or length) $m$, and the random hash functions $\hashdegree(x)$ and  $\hash(x)$. 
An \ac{IBLT} supports several operations: initialization, insertion, deletion, and recovery:
\begin{itemize}
	\item Initialize$()$. This operation sets the different fields  of all the cells in the \ac{IBLT} to zero.
	\item Insert$(\pair)$. The insertion operation \emph{adds} the key-value pair $\pair$ to the \ac{IBLT} (see Algorithm~\ref{alg:insert}).
	\item Delete$(\pair)$. The deletion operation \emph{removes} the key-value pair $\pair$ to the \ac{IBLT} (see Algorithm~\ref{alg:deletion}).
	\item Recover$()$. This operation aims at outputting  all the key-value pairs stored in the \ac{IBLT}. If this operation provides the full list of key-value pairs  in the \ac{IBLT}, we say it succeeded. Otherwise, if it provides an incomplete list, we say the list operation fails (see also Algorithm~\ref{alg:list}).
\end{itemize}
\begin{figure}

\begin{minipage}{0.475\textwidth}
\begin{algorithm}[H]
	\caption{Initialization}\label{alg:init}
	\begin{algorithmic}
		\vspace{-1.5mm}
		\Procedure{Initialize$()$}{}
		\For {i = $1,2,\dots, m$} 
		\State $\cell_{ { i } }.\countfield =0$
		\State $\cell_{ { i } }.\payload =\bm{0}$
		
		\EndFor
		\EndProcedure
	\end{algorithmic}
\end{algorithm}
\end{minipage}\hfill

\begin{minipage}{0.475\textwidth}
\begin{algorithm}[H]
	\caption{Insertion}\label{alg:insert}
	\begin{algorithmic}[H]
		\vspace{-1.5mm}
		\Procedure{Insert$(z)$}{}
		\State $d \gets \hashdegree(z.x)$
		\State $ \jv \gets \hash(z.x)$
		\For {i = $1,2,\dots, d$} 
		\State $\cell_{ \jvi{ i } }.\countfield = \cell_{ \jvi{ i } }.\countfield +1$
		\State $\cell_{ \jvi{ i } }.\payload =\xor \left( \cell_{ \jvi{ i } }. \payload, \,  \pair \right)$
		\EndFor
		\EndProcedure
	\end{algorithmic}
\end{algorithm}
\end{minipage}\hfill
\begin{minipage}{0.475\textwidth}
\begin{algorithm}[H]
	\caption{Deletion}\label{alg:deletion}
	\begin{algorithmic}
		\vspace{-1.5mm}
		\Procedure{Delete$(z)$}{}
		\State $d \gets \hashdegree(z.x)$
		\State $ \jv \gets \hash(z.x)$
		\For {i = $1,2,\dots, d$} 
		\State $\cell_{ \jvi{ i } }.\countfield = \cell_{ \jvi{ i } }.\countfield -1$
		\State $\cell_{ \jvi{ i } }.\payload =\xor \left( \cell_{ \jvi{ i }}.\payload,z\right)$
		\EndFor
		\EndProcedure
	\end{algorithmic}
\end{algorithm}
\end{minipage}

\begin{minipage}{0.475\textwidth}
\begin{algorithm}[H]
	\caption{Recovery}\label{alg:list}
	\begin{algorithmic}
		\vspace{-1.5mm}
		\Procedure{Recover$()$}{}
		
		\While {$\exists i \in [1, m] | \cell_{i}.\countfield = 1  $} 
		\State add $\pair=\cell_{i}.\payload$ to the output list
		\State call Delete $(z)$
		\EndWhile
		\EndProcedure
	\end{algorithmic}
\end{algorithm}
\end{minipage}
\vspace{-4mm}
\end{figure}

\vspace{-2mm}
\subsection{Encoding $\S$ into an IBLT}
The mapping of the $n$ elements of $\S$ to an \ac{IBLT}, also referred to as encoding is done as follows. First, all cells are initialized to zero as described by Algorithm~\ref{alg:init}. After initialization, the elements of $\S$ are successively inserted  into the \ac{IBLT} as described by Algorithm~\ref{alg:insert}: for every  element $\pair=(x,y)$, $d=\hashdegree(x)$ cells with indices $\hash(x)=\jv$ are selected. The element $\pair$ is then \xor-ed with the  \payload field of the cells, and their \countfield field is increased by one.

\subsection{Recovery of $\S$}
We are interested in recovering all $n$ elements of $\S$ from the irregular \ac{IBLT} of length $m$. This process is also referred to as recovery and or decoding. Recovery succeeds if all $m$ cells of the \ac{IBLT} have  \countfield field  equal to zero. In this case the output of the recovery operation will contain all $n$ elements that had been inserted. Otherwise, if some cells have a non-zero count,  recovery fails. 
A low-complexity algorithm  for the recovery of \acp{IBLT} was proposed in \cite{goodrich:2011}, instantiated for a regular \ac{IBLT}. Algorithm~\ref{alg:list} describes the recovery operation for an irregular \ac{IBLT}.  We seek for cells with counter field equal to one, since the  \payload field of such cells is an element $\pair$ of $\S$. Then, $\pair=(x,y)$ is deleted from the \ac{IBLT} by calling Delete$(\pair)$, which removes $\pair$ from 
$\hashdegree(x)=d$ cells with indices $\hash(x)=\jv$. 
Since successful recovery requires processing all $n$ elements of $\S$, and each element gets mapped in average to $\bar{d}$ different cells, complexity of the recovery operations scales as $\mathcal{O}(n \bar{d})$ (which is the same as the encoding complexity).

\subsection{Peeling decoding}
We argue that the recovery operation is an instance of \textit{peeling decoding} \cite{luby1998analysis}. We may represent an \ac{IBLT} as a bipartite (or Tanner) graph $\graph=(\pairs \cup \cells, \edges)$ composed of a set of $n$ data nodes $\pairs$, a set of $m$ cell nodes $\cells$ and a set of edges $\edges$. As the names indicate, data nodes represent key-value pairs and cell nodes represent cells of the \ac{IBLT}. A data node $\pairnode_i \in \pairs$ and a cell node $\cellnode_h \in \cells$ are connected by an edge if and only if  $\pair_i=(x_i,y_i)$ is written to cell $\cell_h$, i.e., $\exists k | \jvi{k} =h$, where $\jv = \hash(x_i) $ and $d = \hashdegree(x_i)$.
A  data node $\pairnode$ and a cell node $\cellnode$ are said to be neighbors if they are connected by an edge. We use the shorthand $\pairnode \in \mathcal{N}(\cellnode)$ or $\cellnode \in \mathcal{N}(\pairnode)$. The degree of a node is given by the number of edges connected to the node. Thus, the degree of a cell node equals the $\countfield$ field of the cell it represents.

Recovery of $\S$ can be represented as a peeling process on a bipartite graph where the graph is \textit{unknown to the decoder} and is revealed during the decoding process. In particular, whenever a cell node $\cellnode$ of degree one is present, its only neighbor $\mathcal{N}(\cellnode)=\pairnode$ is determined. The key-value pair $\pair$ which is represented by the data node $\pairnode$ is added to the output list. Next, the retrieved key-value pair is removed from  the \ac{IBLT} which translates into the removal of all edges attached to its associated data node. 
This process is repeated until no more cell nodes of degree one are present. At this stage, if all cell nodes are of degree zero, recovery succeeded and all key-value pairs are present in the output list. Otherwise, if some cell nodes of degree larger than zero are present, recovery fails, and the output list will not contain all key-value pairs.

\begin{figure}[t]
	\subfloat[$\ell=0$]{
		\includegraphics[width=0.5\columnwidth]{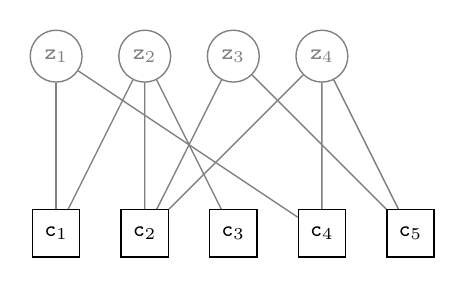}
		
		\label{fig_l_0}
	}
	\subfloat[$\ell=1$]{
		\includegraphics[width=0.5\columnwidth]{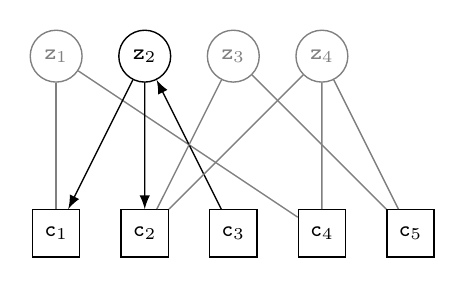}
		\label{fig_l_1}
	}
	
	\subfloat[$\ell=2$]{
		\includegraphics[width=0.5\columnwidth]{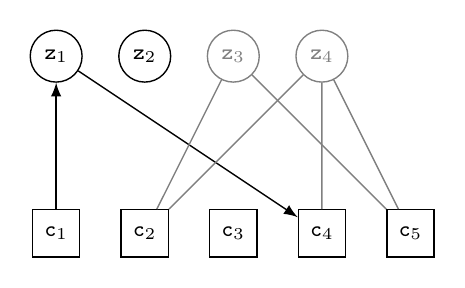}
		\label{fig_l_2}
	}
	\subfloat[$\ell=3$]{
		\includegraphics[width=0.5\columnwidth]{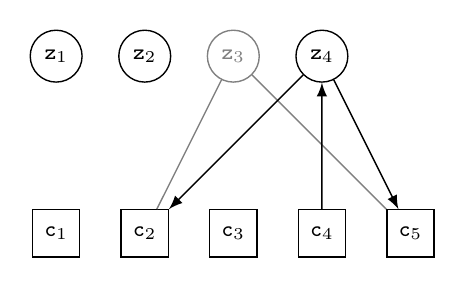}
		\label{fig_l_3}
	}	
	
	\begin{center}
		\subfloat[$\ell=4$]{
			\includegraphics[width=0.5\columnwidth]{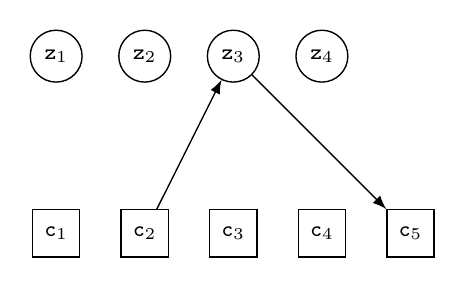}			
			\label{fig_l_4}
		}
	\end{center}
	\caption{Peeling process on the graph representation of an \ac{IBLT} with $n=4$ key-value pairs and $m=5$ cells. The index $\ell$ is used to denote the different steps of the recovery process.}
	\label{fig:graph_peeling}
\end{figure}

\begin{example} [Peeling decoding]
The different steps of the peeling process are shown in Figure~\ref{fig:graph_peeling}. Figure~\ref{fig_l_0} shows the bipartite graph representation of an \ac{IBLT} before the peeling process starts. We observe that  the \ac{IBLT} has  $m=5$ cell and stores  $n=4$  key-value pairs. 
However, at this stage the depicted bipartite graph is \textit{unknown} to the decoder, since it does not have any knowledge about $\S$. The decoder is only aware of the $m$ cell nodes. For this reason, the data nodes as well as the edges are shown in grey. The graph structure will be revealed successively as the recovery operation progresses, and it will only be completely known if decoding succeeds. Otherwise, a part of the graph will remain hidden.
We can  see that cell node $\cellnode_3$ has degree $1$, and thus its associated \ac{IBLT} cell $\cell_3$ has count $1$.  The recovery operation retrieves the only key-value pair that has been mapped to cell $\cell_3$, i.e., data node $\pair_2$, which is added to the output list of the recovery operation. Afterwards, $\pair_2$ is deleted from the \ac{IBLT}. In the graph representation this translates to revealing the only neighbor of cell node $\cellnode_3$, data node $\pairnode_2$ (now shown in black), and deleting all edges attached to it, as shown in Figure~\ref{fig_l_1}. As a consequence, the degree of $\cellnode_1$ becomes one. 
In the next step, as shown in Figure~\ref{fig_l_2}, the only neighbor of $\cellnode_1$, $\pairnode_1$, is revealed and all edges attached to it are removed. This reduces the degree from  $\cellnode_4$ from $2$ to $1$.
Then, data node $\pairnode_4$ is revealed since it is the only neighbor of $\cellnode_4$. After all edges attached to $\pairnode_4$ are removed, as shown in Figure~\ref{fig_l_3},  we have two cell nodes of degree $1$, namely $\cellnode_2$ and $\cellnode_5$, both of which have as only neighbor $\pairnode_3$.
In the last step shown in Figure~\ref{fig_l_4}, first $\pairnode_3$ is revealed as the only neighbor of  $\cellnode_2$. Finally, all edges attached to $\pairnode_3$ are erased from the graph. In this example recovery operation succeeded and set $\S$ was completely recovered.
\end{example}

\vspace{-3mm}
	
	\section{Analysis of the recovery process}
\label{sec:list_analysis}

\subsection{Degree Distributions} \label{sec:dds}

Let us define the node perspective degree distribution polynomial for the data nodes as
\begin{equation}
\Lambda(\x)=  \sum_{d=1}^{\dmax} \Lambda_d \, \x^d
\label{eq:node_pers_pair_}
\end{equation}
where $\x$ is a dummy variable and $\Lambda_d$ is the probability of a data node $\pairnode$ being of degree $d$. 
Similarly,  the node perspective degree distribution polynomial for the cell nodes is
\begin{equation}\label{eq:node_pers_cell}
\Psi(\x)=  \sum_{d=0}^{n} \Psi_d \, \x^d
\end{equation}
where $\Psi_d$ corresponds to the probability of a cell node $\cellnode$ having degree $d$. 
In literature, $\Lambda(x)$ and $\Psi(x)$ are sometimes referred to as \emph{left} and \emph{right} degree distributions, a convention that has its origins in LDPC literature.
It is easy to verify that the average node degrees are obtained by evaluating the derivative of the polynomials in $\x=1$, i.e., $\Lambda'(1)$, and $\Psi'(1)$ respectively.

Note that the data node degree distribution $\Lambda(\x)$ is a design parameter while the cell  node distribution $\Psi(\x)$  is induced by $\Lambda(\x)$, the number of cells $m$ and the cardinality  $n$  of $\S$. In particular, observe that the number of edges connected to the $n$ data nodes must be the same as the number of those connected the $m$ cell nodes, i.e.,
\[
n \Lambda'(1) = m  \Psi'(1) .
\]
Since  $d=\hashdegree(x)$ follows the probability distribution ${P(d=i)=\Lambda_i}$, and  $\jv=\hash(x)$ is a length-$d$ vector of different numbers between $1$ and $m$ chosen uniformly at random without replacement, the probability that a data node $\pairnode$  is connected to a given cell node $\cellnode$ is
\[
\Prob \{ \cellnode \in \mathcal{N}(\pairnode)  \}= \frac{\Psi'(1)}{n}.
\]
If we assume that the outputs of the hash functions   $\hashdegree(x)$ and $\hash(x)$ are independent for different inputs, then the probability that a  cell node $\cellnode$ is connected to $d$  data nodes follows a binomial distribution,
\[
\Psi_d= \binom{n}{d} \left( \frac{\Psi'(1)}{n}\right)^d  \left( 1- \frac{\Psi'(1)}{n}\right)^{n-d}.
\]

Instead of the node-perspective degree distributions, one may also use edge-perspective degree distributions. Let us define by $\lambda_d$ (and $\rho_d$) the probability that a generic edge in the bipartite graph is connected to a degree $d$ data node (a degree $d$ cell node). We have
\[
\lambda_d = \frac{\Lambda_d d}{\sum_\ell \Lambda_\ell \ell} \qquad \text{and} \qquad  \rho_d = \frac{\Psi_d d}{\sum_\ell \Psi_\ell \ell}.
\]
For convenience, the polynomial representations of $\lambda_d$ and $\rho_d$ are chosen to be
\[
\lambda(\x) = \sum_{d} \lambda_d \x^{d-1}
\qquad \text{and} \qquad
\rho(\x) = \sum_{d} \rho_d \x^{d-1}.
\] 

\subsection{Density Evolution}
\label{sec:asymp_anal}
Let us define the  \textit{load} $\eff=n/m$ as the ratio between the number of key-value pairs and cells, and let us consider the regime in which $n$ and $m$ tend to infinity while keeping the load $\eff$ constant.
For a given $\lambda(\x)$ and load  $\eff$  we are interested in determining whether the recovery operation will be successful or not. 
In literature, the performance of peeling decoding is analyzed via \textit{density evolution} \cite{luby1998analysis,richardson2001design}, which restates the peeling decoder as an \textit{equivalent} iterative message passing algorithm where nodes pass messages along the edges to their neighbors. In our case, the messages exchanged by the nodes can be either an \textit{erasure}, i.e., we do not know the corresponding key-value pair yet, or the opposite, \textit{non-erasure}, meaning that key-value pair has been recovered.
In particular, given an ensemble of bipartite graphs $\graphensemble$  with $n$ data nodes, $m=\eff/n$ cell nodes, and edge oriented degree distribution $\lambda(\x)$, density evolution yields the average probability of the exchanged messages at the $\ell$th iteration being an erasure assuming that $n$ goes to infinity.

Denote by $p_\ell$ the (average) probability that the message sent from a cell node over an edge at the $\ell$th iteration is an erasure and by $q_\ell$ the (average) probability that the message sent from a data node over an edge at the $\ell$th iteration is an erasure. 
Consider first the message sent by a cell node of degree $d$ over a given edge. This message will  be a non-erasure if the messages received through the remaining $d-1$ edges were non-erasure messages. Thus we have 
\begin{align}
	1- p_\ell &= (1-q_{\ell})^{d-1} \\
	p_\ell &= 1- (1-q_{\ell})^{d-1} .
\end{align}
Similarly, if we consider a data node of degree $d$, the message sent over an edge will be an erasure only if all messages received over all other $d-1$ edges were erasures. 
Thus,
\[
q_\ell = {p_{\ell-1}}^{d-1} .
\]
We are interested in the average erasure probability, where the average is taken over all edges of all bipartite graphs in $\graphensemble$, hence we have
\begin{align}
	q_\ell = \sum_d \lambda_d {p_{\ell-1}}^{d-1} = \lambda(p_{\ell-1}) .
	\label{eq:q_ell}
\end{align}
Similarly, the average probability that a message sent  by a cell node over a random edge is an erasure can be obtained as
\begin{align}
	p_\ell = \sum_d \rho_d \left( 1- (1-q_{\ell})^{d-1} \right)
	= 1 - \rho(1-q_{\ell}).
	\label{eq:p_ell}
\end{align}
Initially, we have $q_0=p_0=1$, i.e., we start by setting all messages to erasures. Then, by iteratively applying \eqref{eq:q_ell} and \eqref{eq:p_ell} we can track the evolution of $q_\ell$ and $p_\ell$ as the number of iterations $\ell$ grows.
Note that $q_\ell$ corresponds to the probability that a randomly chosen key-value pair has been recovered after $\ell$ iterations.

As shown in \cite{luby1998analysis},  the probability of non-erasure (i.e., success) is subject to a threshold effect (or phase transition) at $\eff=\eff^*$, referred to as load threshold in the sequel. In particular, the list operation will be successful with probability tending to $1$ for loads $\eff$ fulfilling  $0< \eff \leq \eff^*$.
According to \cite{luby1998analysis}, the load threshold $\eff^*$ can be formally expressed as the maximum value of $\eff$ for which 
\begin{equation}
q > \lambda(1 - \rho(1-q) ), \qquad \forall q \in (0,1] .
\label{eq:thres_cond}
\end{equation}
Note that the dependency on $\eff$ is implicit in $\rho(\x)$. In particular, in the asymptotic regime when $n\rightarrow \infty$, we can express $\Psi(\x)$ as
\begin{equation}
\Psi(\x) = e^{-\Psi'(1)  (1-\x)} = e^{-\eff \, \Lambda'(1) \, (1-\x)  }	
\end{equation}
which allows to rewrite $\rho(\x)$ as
\begin{equation}
\rho(\x) = \frac{\Psi'(\x)}{\Psi'(1)} = e^{-\eff \, \Lambda'(1) \, (1-\x) }.
\label{eq:rho_asympt}
\end{equation}
Substituting $\rho(\x)$ in  \eqref{eq:thres_cond} by \eqref{eq:rho_asympt} yields
\begin{equation}
q > \lambda  \left( 1 - e^{ \eff \Lambda'(1) q }\right) , \qquad \forall q \in (0,1]
\label{eq:thres_iblt}
\end{equation}
which explicitly shows the dependency on $\eff$.

	\subsection{Connection to IRSA}
For the bipartite graphs used to represent \acp{IBLT} the left degree distribution $\Lambda(\x)$ (or $\lambda(\x)$) is a free parameter whereas the right degree distribution $\Psi(\x)$ corresponds to a binomial distribution.
Such bipartite graphs have been studied in depth in the context of a random access protocol known as \ac{IRSA} over the collision channel \cite{Liva2011}. A few  important results on such graphs are listed in the following.
The asymptotic regime was first studied in \cite{Liva2011}, where a density evolution analysis was presented.
In \cite{narayanan2012iterative} a sequence of capacity achieving degree distributions was presented, i.e., ensembles whose load threshold converges to $\eff^*=1$.
For the finite length regime, 
an approximate error-floor analysis was presented in \cite{ivanov_2015_finite} whereas an approximate analysis of the waterfall performance was presented in \cite{graell2018waterfall}.

	\section{Numerical Results}

Table~\ref{table:thres} shows the load thresholds  $\eff^*$ for different regular and irregular data node degree distributions obtained via density evolution. For regular distributions, we observe that the load thresholds obtained with the analysis in Section~\ref{sec:list_analysis} coincide with the thresholds reported in \cite{goodrich:2011}, where a different technique was used to obtain the thresholds.\footnote{In \cite{goodrich:2011} results are reported in terms of $1/\eff$.} Among the regular distributions,  $\Lambda(\x)=\x^3$ yields the best threshold $\eff^*=0.8183$.

In addition to regular distributions, Table~\ref{table:thres} also provides the load thresholds for three irregular distributions whose load thresholds are higher than those of regular distributions. 
The slightly irregular distribution $0.887\x^3 + 0.113\x^{21}$ with threshold $0.92$ is taken from \cite{rink2013mixed}, where it was conjectured to be the best irregular distribution with two degrees. 
The distribution $0.25\x^2+0.6\x^3+0.15\x^8$ for \ac{IRSA} is taken from \cite{Liva2011}, and was designed to exhibit good performance for moderate values of $m$. 
Additionally, following the analysis in Section~\ref{sec:list_analysis} we derive the degree distribution $0.15\x^2+0.725\x^3+0.125\x^{18}$ with threshold $0.934$ by using an optimization algorithm called simulated annealing. In particular, the goal of the optimization was maximizing the load threshold, see \eqref{eq:thres_cond}, while limiting the probability of degree 2 since it is associated with high error floors for small and moderate values of $m$ \cite{ivanov_2015_finite}.
\begin{table}
\caption{Load thresholds $\eff^*$ for different degree distributions}
\label{table:thres}
\centering
\begin{tabular}{@{}rl@{}} \toprule 
	$\Lambda(x)$ & $\eff^*$ \\ \midrule
	$\x^3$ &   0.818    \\ 
	$\x^4$ &   0.772    \\
	$0.887\x^3 + 0.113\x^{21}$ & 0.920\\
	$0.25\x^2+0.6\x^3+0.15\x^8$ & 0.892 \\
	$0.15\x^2+0.725\x^3+0.125\x^{18}$ &  0.934\\	
	\bottomrule	 
\end{tabular}
\vspace{-3mm}
\end{table}

\begin{figure}
	\includegraphics[width=0.5\textwidth]{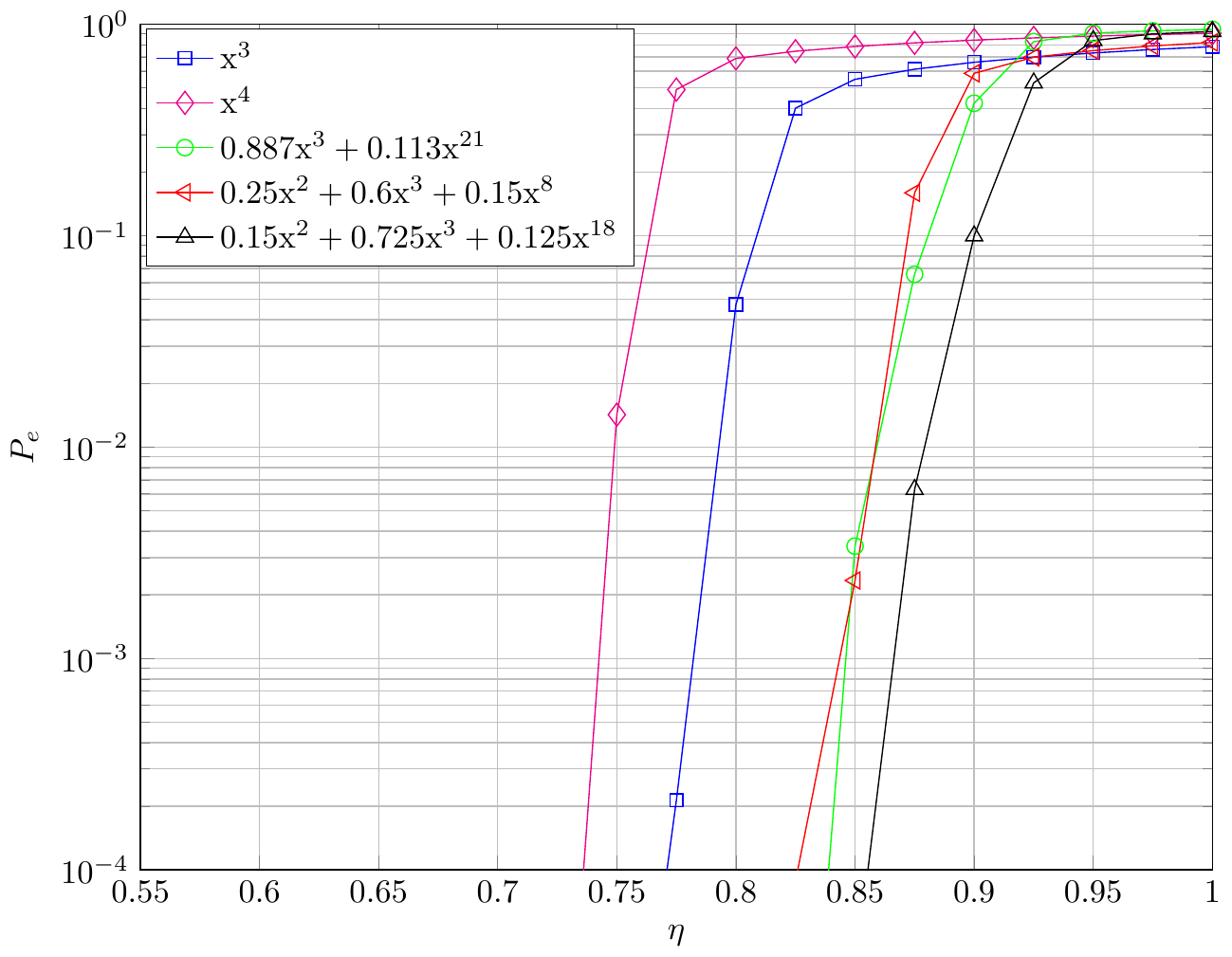}
	\caption{Probability of unsuccessful of a  key-value pair, $P_e$, as a function of $\eff$ for different regular and irregular data node degree distributions with $m=2000$.}
	\label{fig:results}
\end{figure}

Monte Carlo simulations to determine the probability of a key-value pair not being recovered (not present in the output list of the recovery operation), termed $P_e$, versus the channel load $\eff$ are shown in Figure~\ref{fig:results}. We simulated \acp{IBLT} with $m=2000$ for the different degree distribution in Table~\ref{table:thres}.
If we compare the curves in the figure with the asymptotic load thresholds in Table~\ref{table:thres}, we observe that the load threshold provides a good estimate of the load for which $P_e$ undergoes a phase transition, i.e., for which $P_e$ shows a sharp drop. 
The irregular distributions outperform their regular counterparts. 
Among the presented distributions,  $\Lambda(\x)=0.15\x^2+0.725\x^3+0.125\x^{18}$ found by simulated annealing yields the best performance.

\section{Conclusion and Outlook}\label{sec:conclusions}

In this paper we discuss degree distributions for irregular \acp{IBLT}. Realizing recovery corresponds to peeling decoding, we provide a density evolution analysis, which is a novel tool to analyze \acp{IBLT} and extends results from the literature. Furthermore, we show that the graphs induced by \acp{IBLT}, are characterized by a binomial right degree distribution, a family of graphs which has been studied in the framework of random access protocols. This allows to borrow powerful tools from the literature for future work on \acp{IBLT}. Finally, using density evolution we design a degree distribution which  outperforms known distributions for \acp{IBLT}.

Despite the fact that the bipartite graphs arising from \acp{IBLT} have been studied in practice in the context of \ac{IRSA}, some questions related to \acp{IBLT} still remain open. First, in the context of \ac{IRSA} the interesting regime is that of small or moderate values of $m$, due to latency constraints. 
Also, owing to energy efficiency considerations, \ac{IRSA}  distributions usually feature a low average and maximum degree. For \acp{IBLT} scenarios with larger $m$, larger average and maximum degrees might be of interest. Second, and more importantly, 
for some applications \cite{eppstein:2011}, at the time in which the size of the \ac{IBLT} is fixed, the number of key-value pairs which will be inserted in it is not known or deviates strongly from its estimated value.
So far, schemes based on \acp{IBLT} solve this by oversizing the \ac{IBLT}, i.e. operating at lower loads, which is inefficient. A more advantageous scheme would be one that allows to add \ac{IBLT} cells on demand, similarly as it is done in frameless ALOHA \cite{lazaro2020frameless}.
	
	\section*{Acknowledgements}
	The authors would like to thank Federico Clazzer for providing the software used for the Monte Carlo simulations.



\end{document}